# 3D Corporate Tourism in the Marine Sciences: Application-Oriented Problem Solving in Marine and Coastal Ecosystems

ILLE CHRISTINE GEBESHUBER[1,2], TINA REZAIE MATIN[1], RANEE ESICHAIKUL[3], MARK MACQUEEN[4] & BURHANUDDIN YEOP MAJLIS[1]

[1]Institute of Microengineering and Nanoelectronics, Universiti Kebangsaan Malaysia, 43600, Bangi, Selangor, Malaysia
[2]Institute for Applied Physics, Vienna University of Technology, 1040 Wien, Austria
[3]School of Management Science, Sukhothai Thammathirat Open University, Nonthaburi 11120, Thailand
[4]Aramis Technologies, 57000 Kuala Lumpur, Malaysia

**Running Title:**   3D Tourism

**Corresponding Author:**   Prof. Dr. Ille C. Gebeshuber
Institute of Microengineering and Nanoelectronics
Universiti Kebansaan Malaysia
43600 UKM, Bangi

M +60 12 392 9233
T  +60 3 8921 6305
F  +60 3 8925 0439
E  gebeshuber@iap.tuwien.ac.at,
   ille.gebeshuber@mac.com
H  http://www.ille.com
   http://www.ukm.my/imen

# 3D Corporate Tourism in the Marine Sciences: Application-Oriented Problem Solving in Marine and Coastal Ecosystems

">
ILLE CHRISTINE GEBESHUBER[1,2], TINA REZAIE MATIN[1], RANEE ESICHAIKUL[3], MARK MACQUEEN[4] & BURHANUDDIN YEOP MAJLIS[1]

[1]Institute of Microengineering and Nanoelectronics, Universiti Kebangsaan Malaysia, 43600, Bangi, Selangor, Malaysia
[2]Institute for Applied Physics, Vienna University of Technology, 1040 Wien, Austria
[3]School of Management Science, Sukhothai Thammathirat Open University, Nonthaburi 11120, Thailand
[4]Aramis Technologies, 57000 Kuala Lumpur, Malaysia

ABSTRACT

3D corporate tourism in the marine sciences is a solution-based approach to innovation in science, engineering and design. Corporate international scientists, engineers and designers work with local experts in Malaysian marine and coastal environments: they jointly discover, *develop* and **design** complex materials and designs inspired by nature directly on site (e.g. at the UKM Marine Ecosystem Research Centre EKOMAR and Malaysian Marine Parks) and construct initial biomimetic prototypes and novel designs. Thereby, new links, networks and collaborations are established between communities of thinkers in different countries. 3D tourism aims at mapping new frontiers in emerging engineering and design fields. This provides a novel way to foster and promote innovative thinking in the sciences, and considers the need for synergy and collaboration between marine sciences, engineering and design rather than segmentation and isolation.

With the concept of 3D corporate tourism the potential of Malaysian marine ecosystems is used in a sustainable way and the management of marine resources for human and environmental wellbeing is fostered, without exploiting the natural resources or removing anything else from the ecosystem apart from ideas. Subsequent deeper and more detailed investigations at the respective international home institution foster collaborations and result in synergistic effects across borders.

## INTRODUCTION

In Malaysia, tourism is the second largest foreign exchange earner. In addition, it is a multi-sectored industry that consists of transportation, accommodation, restaurants, recreation, entertainment, retailers, handicraft and tour agencies. In 2007, the tourism industry provided employment for almost one million people in Malaysia. Tourism also provides a platform for realising socio-economic and distributive benefit policies. Community-based tourism principles are applied in the implementation of homestay and eco-tourism programmes. Community-based tourism strengthens the ability of rural communities to manage tourism resources, earn their own income while ensuring local participation (Peng, 2009).

3Niche Tourism refers to special interest tourism (Macleod, 2003) and is a way forward to sustainability (Novelli and Benson, 2005). Malaysia is increasingly focusing on niche tourism markets (The Star, 2008; The Star, 2010).

Policies aiming at innovation in marine sciences should not only focus on the sciences itself, but take into account the driving forces of other business sectors and the public sector. This manuscript introduces a niche tourism concept where the marine sciences, engineering and design are the driving forces for innovation in tourism: 3D corporate tourism (for logo, see Figure 1), a solution–based approach to innovation in science, engineering and design (Gebeshuber and Majlis, 2010a; Menon et al., 2010). The three main pillars of this integrated concept are termed discover, *determine* and **design** (Figure 2). In 3D corporate tourism for the marine sciences, corporate international biologists, scientists, engineers and designers jointly work with Malaysian marine scientists and locals in marine and coastal environments (high inspirational potential!) and construct initial prototypes and designs directly on site. This joint niche tourism approach yields new links, networks and collaborations between communities of thinkers and local populations (including indigenous peoples and their tacit knowledge) in different countries in order to stimulate and enhance creative and application–oriented problem solving for society.

The high species variety in marine ecosystems such as coastal seas, lagoons, estuaries, mangroves, coral reefs and deep oceanic waters, with nature's 'best practices' everywhere aids to relate structure with function in natural materials, structures and processes and helps to increase awareness about the natural resources surrounding us.

With the concept of 3D corporate tourism the potential of Malaysian marine ecosystems is used in a sustainable way and the management of marine resources for human and environmental wellbeing is fostered, without exploiting the natural resources or removing anything else from the ecosystem apart from ideas. In this way, the value of marine ecosystems is increasing in the minds of policy makers and threshold countries have the opportunity to contribute highly valued inputs to the international research and development elite, as well as train their local experts in important future technologies. The possibility to perform first investigations directly on–site (e.g. at the UKM Marine Ecosystem Research Centre EKOMAR and Malaysian marine parks), and subsequent deeper and more detailed investigations at the home institution fosters collaborations and results in synergistic effects across borders.

MATERIALS AND METHODS

With 3D corporate tourism in the marine sciences, the successful concept of the 'Biomimicry and Design Workshops' offered by the US based Biomimicry Guild is developed further into a complete niche tourism concept. At the 'Biomimicry and Design Workshops', researchers and corporate members spend about one week in the rainforest, and perform the Biomimicry Innovation Method (see below) with the help of specifically trained biologists (called 'biologists at the design table). Databases and scientific articles are available on local hard-drives, and access to the Internet is provided on occasional visits to local internet cafés.

The Biomimicry Innovation Method (Biomimicry Guild, 2008) is an innovation method that seeks sustainable solutions by emulating nature's time-tested patterns and strategies. The goal is to create products, processes, and policies - new ways of living - that are well adapted to life on earth over the long haul. The Biomimicry Innovation Method involves specifically trained biologists as well as engineers, natural scientists, architects and/or designers from universities or companies. This method is for example applied in the rainforests or in corral reefs to learn from and emulate natural models.

The steps in the Biomimicry Innovation Method are identify function, biologize the question, find nature's best practices and generate product ideas. *Identify function:* The marine scientists distil challenges posed by engineers/natural scientists/architects and/or designers to their functional essence. *Biologize the question:* In the next step, these functions are translated into biological questions such as "How are vibrations amplified or attenuated in marine organisms?" or "How do marine organisms direct vibrations?" or "How do marine organisms manage high impacts?" The basic question is "What would marine organisms or systems do here?" *Find nature's best practices:* Screens of the relevant literature in scientific databases as well as entering highly inspiring environments with the biologized questions in mind (task-oriented visit) are used to obtain a compendium of how marine or coastal plants, animals and systems solve the challenges in question. The inspiring environments should preferably be habitats with high species diversity, e.g., coastal seas, lagoons, mangrove forests and coral reefs. Thereby a compendium of how marine or coastal plants, animals and ecosystems solve the specific challenge is obtained. *Generate process/product ideas:* From these best practices ideas for new designs, materials, structures or processes are generated.

The three Ds in 3D corporate tourism stands for the three pillars of this concept: discover, *determine* and **design** (Figures 2 and 3). In the discover phase, problems (in science, in technology, in design) are formulated and respective solutions in marine and coastal environments are sought with the help of nature scouts and Malaysian marine scientists (Figures 4 and 5). Nature scouts are locals and/or people from indigenous communities, who know their surroundings and understand the complex biosystems of their environment. The next phase is the *determine* phase. Here, details are analyzed and first designs of solutions are drafted. Process experts analyse such solutions regarding energy balance, costs and benefits. The best method for choosing fit solutions and for the design is selected via a push-pull analysis. This push-pull analysis investigates the needs of the corporate specialists in relation to the available potential in nature (at reasonable cost). Pull: There is pull from the corporate specialists, defining their requirements. Push: The available solutions in nature for the creation of the man-made solution will be assessed. In the **design** phase, nature's solutions are adapted to human technology and design; nature serves as design study.

3D corporate tourism aims at mapping new frontiers in emerging and developing engineering and design areas. It provides a novel way to foster and promote innovative thinking in the sciences, and considers the need for synergy and collaboration between marine sciences, biology, engineering and materials science rather than segmentation and isolation: Corporate specialists (i.e., specifically trained biologists, scientists, engineers as well as designers) travel to adequate places (i.e., marine and coastal environments with high inspirational potential) in Malaysia and apply the Biomimicry Innovation Method that promotes knowledge transfer from nature to technology development and design (biomimetics; Bar-Cohen, 2005; Gleich, Pade, Petschow and Pissarskoi, 2010). The international corporate specialists are supported by Malaysian marine scientists (Figures 4 and 5) and discover, *determine* and **design** complex materials and design solutions inspired by nature. Directly at the site of this research, first prototypes and designs are constructed, and first detailed investigations take place. Access to the Internet and to databases is continually provided, and machinery to investigate structures and materials on-site is provided. In contrast to the 'Biomimicry and Design Workshops' the research stations where the corporate specialists are staying are equipped with all amenities that executives might need, and with specific libraries and CAD and further prototyping machines. To facilitate networking, discussion and creativity, locals and corporate specialists stay together for the whole duration of the stay. After the first initial prototypes are constructed, a family holiday for the corporate specialists takes place, freeing the mind and allowing for new ideas and concepts to settle. After the family holiday, the corporate specialist and their local partners meet again, to finalize the designs and draft future joint projects and collaborations.

Nature can serve as teacher for technology and design development (Gebeshuber, Stachelberger and Drack, 2005; Gebeshuber and Crawford, 2006; Srajer, Majlis and Gebeshuber, 2009), since nature's materials and structures are complex, multi–functional, hierarchical and responsive, and in most cases far better than man-made materials. Increasingly, collaborations across fields (such as among marine scientists, the hard sciences, engineering, design, art and the knowledge of local and indigenous communities) prove successful (Gebeshuber, 2007; Gebeshuber and Drack, 2008; Gebeshuber et al., 2010; Bawitsch and Stemeseder, 2010) and are highly useful for innovation, e.g. in naoscience and nanotechnology (Gebeshuber, Gruber and Drack, 2009; Gebeshuber et al., 2009; Gebeshuber, Majlis and Stachelberger, 2009).

The outcome of 3D corporate tourism are – besides the research results, developments and designs – new links, networks and collaborations between communities of thinkers in different countries in order to stimulate and enhance creative and application–oriented problem solving for society (Figure 4).

RESULTS

3D corporate tourism in the marine sciences is a high quality niche tourism area that sustainably applies and expands the Biomimicry Innovation Method. Preliminary data on how biomimicry can contribute to niche tourism were acquired on a scientific expedition to Bukit Fraser in February 2010 (organizer: Prof. Jumaat Adam, Universiti Kebangsaan Malaysia).

The corporate executives were two tissue engineers from Austria (Jennifer Bawitsch and Teresa Stemeseder), performing industry training at the Institute of Microengineering and Nanotechnology at Universiti Kebangsaan Malaysia, and one of the authors, ICG, who is a physics professor at UKM and the Vienna University of Technology in Austria, Europe. Prof. Jumaat Adam and his team from the UKM biology department and various locals such as our guide Mr. Shukri from the Malaysian forestry department and local rangers served as bioscouts. Joint publications and reports as well as various ideas for future collaborations and research projects were generated during the stay at the research station in Bukit Fraser (director: Prof. Jumaat Adam). The executives also spent their subsequent holidays in Malaysia, with several additional friends and colleagues from Austria, who were attracted by their reports and success stories (Bawitsch J. & Stemeseder T., 2010).

Thereby, a first proof of principle on how the Biomimicry Innovation Method links to niche tourism and corporate applications of biomimicry design workshops was established. This initial application of the concept of 3D corporate tourism shows how technology, especially in terms of design and innovation, is helpful for tourism-based schemes.

One example for successful biomimetics is the scientific treatment of fish slime, a highly multifunctional material. Proposed roles are in respiration, ionic and osmotic regulation, reproduction, excretion, disease resistance, communication, feeding, nest building and protection (Shepard, 1994). Fish body slime not only helps reduce drag, but also reduces the vibrations caused by passing waves. Rosen and Cornford showed an impressive 66% reduction of friction of a diluted solution of fish slime (Rosen and Conford, 1970). Bechart and co-workers (2000) experimentally investigated wall shear stress reduction in sharkskin replica consisting of 800 plastic model scales with compliant anchoring. Self-cleaning surfaces such as that of lotus leaves represent an interesting option to avoid fluid-dynamic deterioration by the agglomeration of dirt. An example of technological implementation of such as system combing sharkskin and lotus leaf properties are long-range commercial aircraft. Separation control by bird feathers inspires self-activated movable flaps (i.e., artificial bird feathers) that represent a high-lift system enhancing the maximum lift of airfoils by about 20%. Bechart and co-workers show that this is achieved without perceivable deleterious effects under cruise conditions (Bechert, Bruse, Hage, and Meyer, 2000). A fish slime inspired aircrafts outer coating may even help attenuate sound (Gebeshuber and Majlis, 2009).

## DISCUSSION

The high species variety in Malaysian marine and coastal environments, with nature's 'best practices' everywhere aids to relate structure with function in natural materials, structures and processes and helps to increase awareness about the natural resources surrounding us. With the 3D corporate tourism concept in the marine sciences the potential of marine and coastal environments is used in a sustainable way, without exploiting the natural resources or removing anything else from the ecosystem apart from ideas and inspiration. In this way, the worth of such ecosystems is increasing in the minds of policy makers and threshold countries have the opportunity to contribute highly valued inputs to the international research and development elite, as well as train their local experts in important future technologies. The possibility to perform first investigations directly on–site, and subsequent deeper and more detailed investigations at the home institution fosters collaborations and results in synergistic effects across borders, locally and mentally (Gebeshuber and Majlis, 2010b).

3D Nexus is the general connection between local experts, bioscouts, researchers and industrial designers (Figure 5). 3D Nexus stands for a highly profitable Malaysian centre of competence that serves as basis for a completely new type of tourism. In 3D tourism, corporate and scientific tourism merge with ecotourism. The tourists themselves are high-end tourists such as executives and professors for industrial design, ecology, business and the pharmaceutical industry.

The Malaysia marine and coastal environments are perfect for realizing 3D tourism: they offer high biodiversity and well-equipped marine parks. In various marine stations, habitats of interest such as the high sea areas, riffs and mangrove forests as well as coastal areas with industry are at the disposal of the corporate tourists, ensuring the essential diversity of possible solutions in a variety of zones.

Advantages of this concept are an increased appreciation of marine and coastal areas, combined with comprehensive collection and sustainment of the knowledge of indigenous peoples. Furthermore, the input from nature for technology inherently includes best practices and allows for the evaluation of existing solutions in nature according to the secondary impact (technology assessment): also in nature some solutions yield unfavourable consequences. An example for this is the introduction of tilapiine fish species to lake Victoria in Africa in the middle of the last century, which lead to a decline, and in some cases an almost total disappearance, of many of the native fish species of lake Victoria (Ogutu-Ohwayo, 1990).

## SUMMARY AND OUTLOOK

The benefits of realization of the 3D corporate tourism concept for the Malaysian marine scientist community would be tighter embedding in international research networks, international research projects, collaborations with communities of thinkers in different countries and a supplementary form of income from international corporate tourists. Malaysian marine scientists would furthermore benefit from a novel way to foster and promote innovative thinking.

The benefits of realization of this concept for the local community comprises increased funding for enhancement for community structures such as streets and joint facilities. The central government would benefit from ideas for new community projects that increase the attractivity of coastal and marine environment for corporate tourists. Malaysian companies would provide the needed facilities and instruments such as Internet access, CAD machines, microscopes, design materials and information material.

Summing up, 3D corporate tourism in the marine sciences is a promising concept that should be developed further and be implemented in marine science programmes and projects in Malaysia.

## ACKNOWLEDGEMENTS

The Austrian Society for the Advancement of Plant Sciences funded part of this work via the Biomimetics Pilot Project "BioScreen". Living in the tropics and exposure to high species diversity at frequent excursions to marine ecosystems is highly inspirational for doing biomimetics. Profs. F. Aumayr, H. Störi and G. Badurek from the Vienna University of Technology are acknowledged for enabling ICG three years of research in Malaysia.

FIGURE CAPTIONS

Figure 1   Logo of 3D Corporate Tourism

Figure 2   Concept of 3D Corporate Tourism

Figure 3   The 3D Approach

Figure 4   Interactions of the society, local marine experts, industrial developers and marine bioscouts

Figure 5   The central role of the 3D Nexus

FIGURE 1

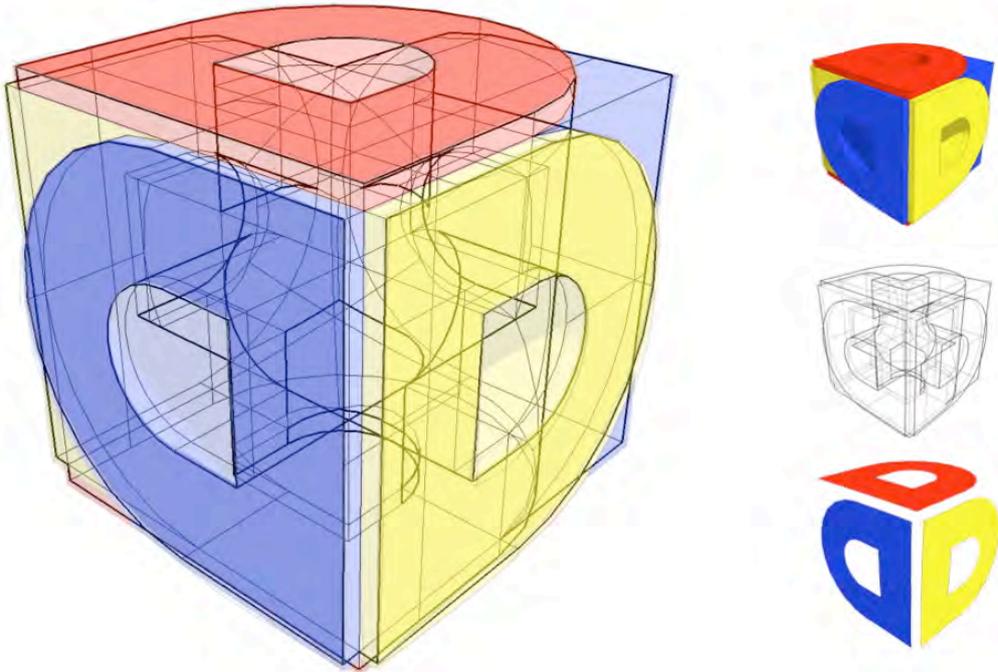

FIGURE 2

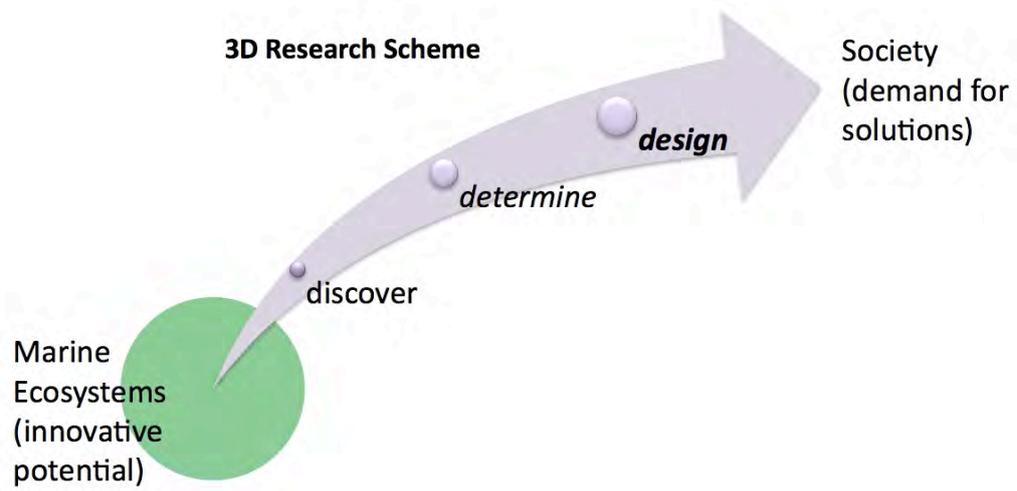

FIGURE 3

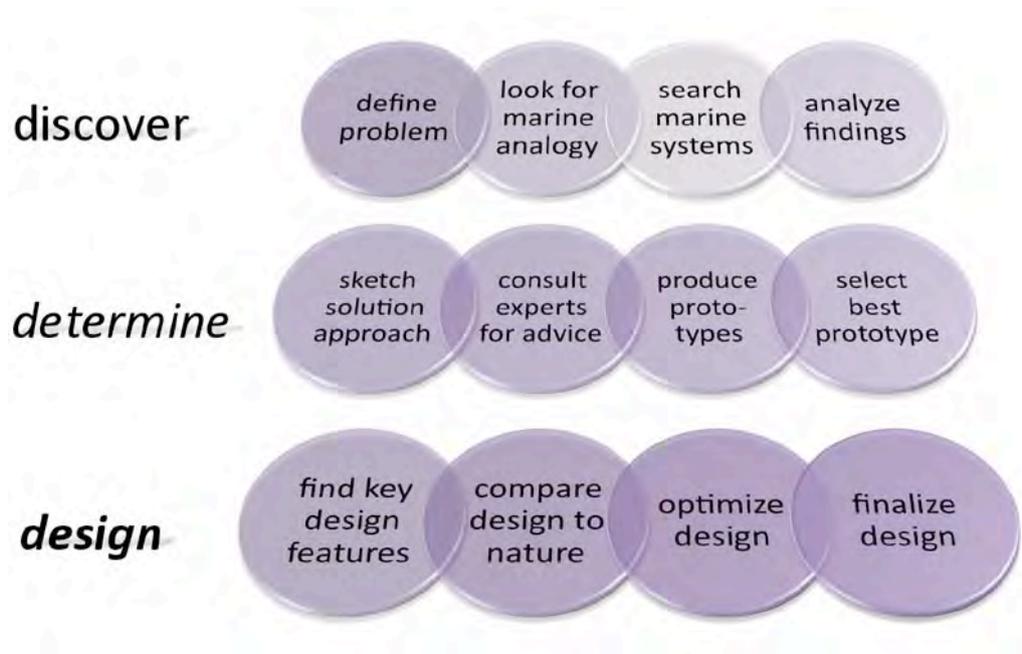

FIGURE 4

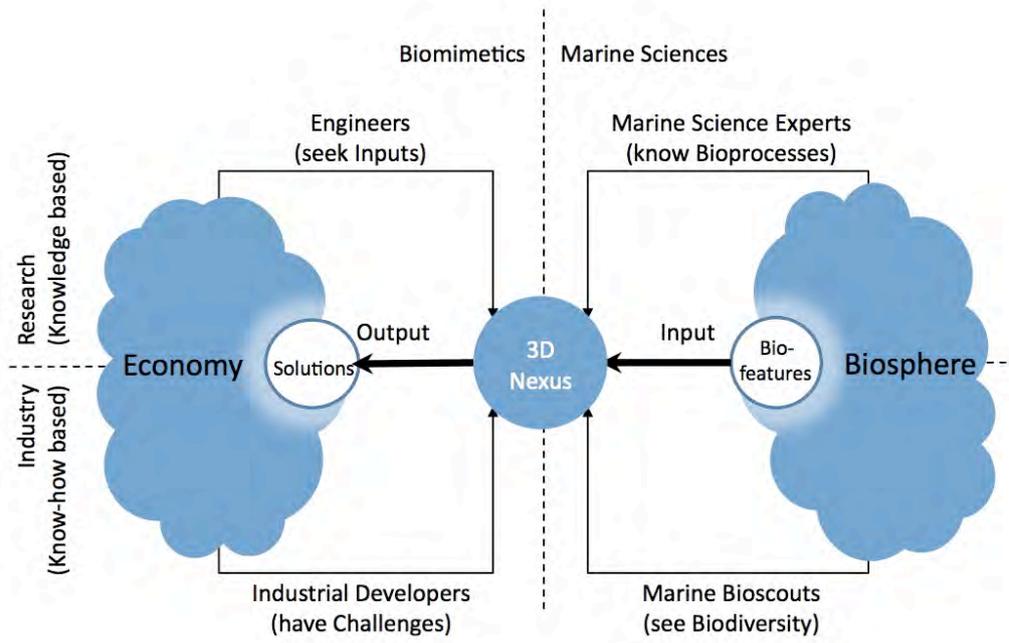

FIGURE 5

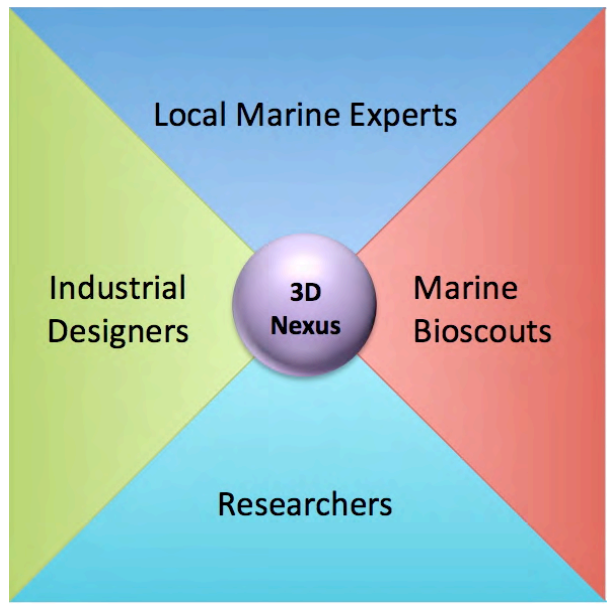